\documentclass[twocolumn,showpacs,preprintnumbers,amsmath,amssymb,aps]{revtex4-1}
\usepackage{graphicx}

\begin{document}

\title{Tight-binding approach to understand photoelectron intensity from graphene for circularly polarized light}

\author{Hwihyeon Hwang}
\author{Choongyu Hwang}
\email[]{ckhwang@pnu.ac.kr}

\affiliation{Department of Physics, Pusan National University, Busan
609-735, Republic of Korea}

\date{\today}

\begin{abstract}
We have investigated the effect of imperfect circular polarization
on the angle-resolved photoemission spectroscopy signal, using
graphene as a prototypical system that can be understood within
tight-binding formalism. We found that perfect left- and
right-circularly polarized lights give the same photoelectron
intensity distribution around a constant energy contour of the
graphene $\pi$ band. On the other hand, upon breaking the purity of
the polarization, photoelectron intensity starts to show circular
dichroism, which is enhanced with further increasing the
imperfection. Our results predict the existence of an additional
factor for the circular dichroism observed in the photoemission
signal from graphene and hence suggest the importance of
experimental conditions to understand circular dichroism observed
via photoemission spectroscopy.
\end{abstract}

\pacs{}

\maketitle

\section{Introduction}

Circular dichroism has been one of the powerful methodologies to
extract information on spin and/or orbital properties of charge
carriers in solid state systems~\cite{Kuch,Polcik,Mulazzi}. In
addition, recent study on graphene has suggested that information on
Berry phase can also be obtained via the circular
dichroism~\cite{Liu}, extending a previous approach of the direct
measurement of Berry phase using linearly polarized lights for the
same system~\cite{Hwang}. These results provide an experimental
evidence that the quantum mechanical phases can be probed by
photoemission spectroscopy, previously not believed to be possible,
and hence constitute the first band specific measurements of Berry
phase.

These interesting observations have been possible due to the simple
geometric structure of graphene, allowing us to obtain the explicit
form of the initial electronic states within the tight-binding
formalism~\cite{Neto}. Shirley {\it et al}.,~\cite{Himpsel}
calculated photoelectron intensity, which is the absolute square of
the transition matrix element $M_{\bf k}=\left<f_{\bf
k}\right|H^{\rm int}({\bf k}) \left|\psi_{\bf k}\right>$, where
$\left|\psi_{\bf k}\right>$ is a tight-binding eigenstate, $\left|
f_{\bf k}\right>$ is a plane-wave final state, and $H^{\rm int}={\bf
A}\cdot{\bf p}$. Here, {\bf A} is a light polarization and ${\bf
p}=-i\hbar\nabla$ is the momentum operator, where $\hbar$ is the
Planck's constant.

This approach reproduces the photoelectron intensity for the
linearly polarized light along the $x$-axis denoted in Fig.~1(a),
i.\,e.\,, X-polarization (${\bf A}=A_x\hat{\bf x}$), whereas it is
not successful in reproducing full polarization dependence of the
photoelectron intensity, e.\,g.\,, when the light polarization is
rotated by 90 degrees, i.\,e.\,, Y-polarization (${\bf
A}=A_y\hat{\bf y}$)~\cite{Hwang}. This issue originates from the
application of ${\bf p}=-i\hbar\nabla$ to the tight-binding
eigenstates. When the tight-binding Hamiltonian is intrinsically
non-local, the derivative in real space for the tight-binding
eigenstate does not work~\cite{Starace,Louie}. This suggests that
the agreement even for X-polarization using {\bf p}~\cite{Himpsel}
could be fortuitous. In order to solve this issue, Hwang {\it et
al}.,~\cite{Hwang} introduced an alternative approach replacing the
derivative to the commutation relation, i.\,e.\,, ${\bf
p}/m_0=-i\hbar\nabla/m_0$, where $m_0$ is free-electron mass, to
${\bf v}=\left[{\bf r},{H}^0\right]/i\hbar$, where ${\bf
r}=i\hbar\,(\nabla_{\bf k},\partial_{k_{\rm z}})$ in the {\bf
k}-representation~\cite{Yu,Louie}.

To avoid this issue, Liu {\it et al.,} have assumed realistic final
states, i.\,e.\,, the Block sum of the Wannier states, while still
using {\bf p}~\cite{Liu}. With this setup, they have claimed that
the observed circular dichroism originates from Berry phase of
graphene, which is challenged by a study on the circular dichroism
as a function of electron binding energy~\cite{Hwang_JPCM}. In fact,
another experimental study shows that the circular dichroism varies
upon changing photon energy, emphasizing the role of the final state
effect in understanding photoemission signal from
graphene~\cite{Gierz}. These controversies give rise to a
fundamental question on the origin of the circular dichroism.

  \begin{figure}[b]
  \begin{center}
  \includegraphics[width=0.95\columnwidth]{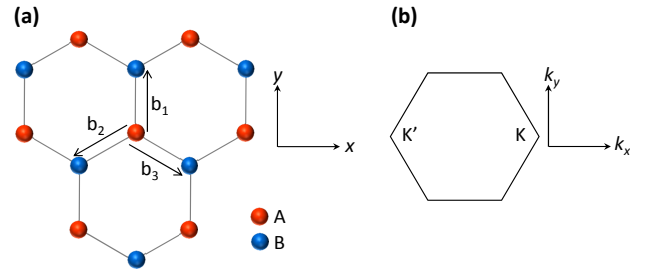}
  \end{center}
  \caption{(a) Schematic of graphene in real space. (b) The first Brillouin zone of graphene. Here, ${\bf
b}_1=b\left(0,1\right)$, ${\bf
b}_2=b\left(-\frac{\sqrt{3}}{2},-\frac{1}{2}\right)$
    and ${\bf b}_3=b\left(\frac{\sqrt{3}}{2},-\frac{1}{2}\right)$ are the three vectors
    connecting the in-plane nearest neighbor atoms
    where $b=1.42$~\AA.
    The positions of the K and K$'$ points are
    $\left(\frac{4\pi}{3a},0\right)$ and
    $\left(-\frac{4\pi}{3a},0\right)$, respectively, where $a=\sqrt{3}b$.}
  \label{Fig1}
  \end{figure}

Here we report calculated photoelectron intensity of graphene for
circularly polarized light. The photoemission matrix element was
constructed using the velocity operator~\cite{Louie,Hwang} within
the tight-binding formalism. We found that the photoelectron
intensity for left- and right-circularly polarized light (LCP and
RCP, respectively) does not show any difference for perfect circular
polarization. On the other hand, upon decreasing the purity of the
polarization, the photoelectron intensity starts to exhibit circular
dichroism. Our results indicate that an experimental condition in
conjunction with the intrinsic effects such as the isospin of charge
carriers and the final states~\cite{Gierz} plays an important role
in determining photoelectron intensity.

\section{Tight-binding formalism}

Figure~1(a) shows the geometric structure of graphene with two
carbon sublattices, A and B. In order to describe the electron
energy eigenvalues and wavefunctions, we have used tight-binding
formalism for the $p_z$ orbital of each sublattice using the
in-plane nearest-neighbor (A-B) hopping integral, $t_0$ (we confine
our interest to this single parameter for simplicity), which
corresponds to $-\gamma_0'$ in the well-known
Slonczewski-Weiss-McClure model~\cite{SWMc1,SWMc2}. Because our
basis set for the tight-binding model has more than one
non-equivalent orbitals, there exist two different hopping
parameters differing only in the sign that give exactly the same
electron band structure. In this sense, the absolute magnitude of
inter-orbital hopping integrals within the empirical tight-binding
Hamiltonian $H_{\rm TB}$ between non-equivalent states (i.\,e.\,,
$\left<\phi_1\right|H_{\rm TB}\left|\phi_2\right>$, where the
localized orbitals $\left|\phi_1\right>$ and $\left|\phi_2\right>$
are not equivalent) in any material have only been speculated
theoretically, whereas it has recently been proved that the sign of
the hopping integrals for both single- and double-layer graphene can
be experimentally determined uniquely using the angle-resolved
photoemission spectroscopy (ARPES) technique~\cite{Hwang}. We have
used $|t_0|=3.16$~eV and $t_0<0$, the values in Table II of
Gr\"uneis {\it et al.,}~\cite{Gruneis} and in Fig.~8 of Hwang {\it
et al}.~\cite{Hwang}, respectively.

The tight-binding Hamiltonian of graphene for two-dimensional
wavevector {\bf k} is as follows~\cite{Hwang,Neto}:
\begin{equation}
H^0({\bf k})=\left(
\begin{array}{cc}
0 & t_0 g({\bf k})\\
\\
t_0\,g^*({\bf k}) & 0\\
\end{array}
\right)\,, \label{eq:H_mono_0}
\end{equation}
using a basis set composed of Bloch sums of localized orbitals on
each sublattice:
\begin{equation}
g({\bf k})=\sum_{i=1}^{3}\exp(i{\bf k}\cdot{\bf b}_i)
\label{eq:g_def}
\end{equation}
with ${\bf b}_i$'s defined as in Fig.~1(a), and
\begin{equation}
\left(
\begin{array}{c}
1\\
0
\end{array}
\right)_{\bf k}
=\frac{1}{\sqrt{N}}\sum_{{\bf R}_{\rm A}}
e^{i{\bf k}\cdot{\bf R}_{\rm A}}\phi({\bf r}-{\bf R}_{\rm A})\,,
\label{eq:1000}
\end{equation}
\begin{equation}
\left(
\begin{array}{c}
0\\
1\\
\end{array}
\right)_{\bf k} =\frac{1}{\sqrt{N}}\sum_{{\bf R}_{\rm B}} e^{i{\bf
k}\cdot{\bf R}_{\rm B}}\phi({\bf r}-{\bf R}_{\rm B})\,.
\label{eq:0100}
\end{equation}

In the presence of the vector potential {\bf A}, the Hamiltonian is
obtained by Peierls substitution, i.\,e.\,, ${\bf k}\to{\bf
k}-\frac{e}{\hbar c}{\bf A}$. Then, the interaction Hamiltonian
$H^{\rm int}$ is obtained by the first-order term of {\bf A} from
$H^0\left({\bf k}-\frac{e}{\hbar c}{\bf A}\right)-H^0({\bf k})$, and
represented as $-\frac{e}{c}\hat{\bf A}\cdot\hat{\bf v}$ using the
velocity operator $\hat{\bf v}=\left[\hat{\bf
r},\hat{H}^0\right]/i\hbar$, where $\hat{\bf r}=i\hbar\,\nabla_{\bf
k}$ in the {\bf k}-representation, $\hbar$ is the Planck's constant,
{\it e} is the charge of an electron, {\it c} is the speed of light.
Then the interaction Hamiltonian becomes~\cite{Hwang}:
\begin{equation}
H^{\rm int}({\bf k})=-\frac{e}{\hbar\,c} {\bf A}\cdot \left(
\begin{array}{cccc}
0 & t_0\,\nabla_{\bf k}g({\bf k})\\
\\
t_0\,\nabla_{\bf k}g^*({\bf k}) & 0\\
\end{array}
\right)\,. \label{eq:H_mono_int}
\end{equation}
Note here that because $g({\bf k})$ depends only on the $k_x$ and
$k_y$ components of the wavevector {\bf k}, there is no contribution
arising from the {\it z} component of the vector potential $A_z$
within this tight-binding model. In real measurements, the light
with a nonzero polarization component along the {\it z} direction
will give rise to an additive isotropic term to the photoelectron
intensity that is independent of the in-plane polarization of the
light.

\section{The analysis of the photoelectron intensity}

The photoelectron intensity is described by the absolute square of
the transition matrix element $M_{s\,{\bf k}}=\left<f_{{\bf
k}}\right|H^{\rm int}({\bf k}) \left|\psi_{s\,{\bf k}}\right>$,
where $\left|\psi_{s\,{\bf k}}\right>$ is graphene eigenstate with
the band index $s=\pm1$ for conduction ($+$) and valence ($-$)
bands, and $\left| f_{{\bf k}}\right>$ is the plane-wave final state
projected onto the $p_z$ orbitals of graphene. For graphene, we may
use
\begin{equation}
\left| f_{\bf k}\right> = \frac{1}{\sqrt{2}}
\left(
\begin{array}{c}
1\\
1
\end{array}
\right)_{\bf k}\,. \label{eq:pw_mono}
\end{equation}
Here, we neglect the $k_z$ dependence of the final state, as done in
the previous works~\cite{Hwang,Himpsel,Eli}. For a few tens eV
photons, typical for ARPES measurements for
graphene~\cite{Liu,Eli,Gierz,Gierz_PRB,Hwang,Hwang_JPCM}, $k_z$ of
the plane-wave final state is much larger than $k_x$ and $k_y$,
leading to only a small variation in $k_z$ with any change in $k_x$
and $k_y$. In addition, the study on $k_z$, i.\,e.\,, photon energy
dependence, is beyond the capability of the tight-binding formalism,
but can be achieved through the first principles
calculations~\cite{Gierz}.

When we consider the case where {\bf k} is very close to the Dirac
point {\bf K} as denoted in Fig.~1(b), and define ${\bf q}={\bf
k}-{\bf K}$ ($|{\bf q}|\ll |{\bf K}|$), Eq.~(\ref{eq:g_def}) becomes
\begin{equation}
g({\bf q}+{\bf K})\approx
-\frac{\sqrt{3}}{2}b\left(q_x-iq_y\right)\,, \label{eq:g}
\end{equation}
and the Hamiltonians become
\begin{equation}
H^0({\bf q}+{\bf K})\approx -\frac{\sqrt{3}}{2}\,b\, t_0 \left(
q_x\,\sigma_x+q_y\,\sigma_y \right)\,, \label{eq:H_mono_0_K}
\end{equation}
and
\begin{equation}
H^{\rm int}({\bf q}+{\bf K})\approx \frac{\sqrt{3}e}{2\hbar c}\,b\,
t_0 \left( {\bf A}\cdot{\bf \sigma} \right)\,,
\label{eq:H_mono_int_K}
\end{equation}
where {\bf $\sigma$} is the Pauli matrix. Then the electron energy
eigenvalues and wavefunctions of the $H^0$ are given by $E_{s\,{\bf
k}}=\frac{\sqrt{3}}{2}\,b\, |t_0|\,s\,|{\bf q}|$ and
\begin{equation}
\left|\psi_{s\,{\bf k}}\right>=\frac{1}{\sqrt{2}}
\left(
\begin{array}{c}
e^{-i\theta_{\bf q}/2}
\\
s\,e^{i\theta_{\bf q}/2}
\end{array}
\right)\,,
\label{eq:wfn_mono}
\end{equation}
respectively, when $\theta_{\bf q}$ is the angle between {\bf q} and
the $+k_x$ direction. With this setup, the photoemission matrix
elements for X- and Y-polarizations are given by
\begin{equation}
M^{\rm X-pol.}_{+1\,{\bf k}}\sim\exp(-i\theta_{\bf
q}/2)+s\exp(\theta_{\bf q}/2), \label{eq:mtx2_mono_2}
\end{equation}
and
\begin{equation}
M^{\rm Y-pol.}_{+1\,{\bf k}}\sim\exp(-i\theta_{\bf
q}/2)-s\exp(\theta_{\bf q}/2), \label{eq:mtx2_mono_2}
\end{equation}
respectively, of which absolute square, i.\,e.\,,
$I_{s\,k}=\left|M_{s\,k}\right|^2$, represents photoelectron
intensity that reproduces the previous experimental results for
linearly polarized lights~\cite{Hwang,Liu}. Resultantly, the
variation of photoelectron intensity depending on the polarity of
light reveals the pseudospin nature of charge carriers in single-
and double-layer graphene, i.\,e.\,, Berry phase of $\pi$ and
$2\pi$, respectively, and the signs of hopping integral in the
tight-binding Hamiltonian for double-layer graphene or
graphite~\cite{Hwang}.

In the same analogy, for the circularly polarized light, i.\,e.\,,
${\bf A}=A_x\,\hat{x}\pm\,i\,A_y\,\hat{y}$, the photoemission matrix
element is given by
\begin{eqnarray}
M_{s\,{\bf k}}&=&\, \frac{A_x}{2}\left(s\,\exp({i\theta_{\bf q}/2})
+\exp(-i\theta_{\bf q}/2)\right)\nonumber\\
&\pm&\,\frac{A_y}{2}\left(s\,\exp({i\theta_{\bf
q}/2})-\exp(-i\theta_{\bf q}/2)\right)\,, \label{eq:mtx_mono}
\end{eqnarray}
where $+$ and $-$ correspond to LCP and RCP, respectively. For the
states above the Dirac energy, i.\,e.\,, $s=+1$
\begin{equation}
M^{\rm LCP}_{+1\,{\bf k}}=\,A_x\cos(\theta_{\bf
q}/2)+i\,A_y\sin(\theta_{\bf q}/2)\,, \label{eq:mtx2_mono_2}
\end{equation}
and
\begin{equation}
M^{\rm RCP}_{+1\,{\bf k}}=\,A_x\cos(\theta_{\bf
q}/2)-i\,A_y\sin(\theta_{\bf q}/2)\,, \label{eq:mtx2_mono_2}
\end{equation}
and for the states below the Dirac energy, i.\,e.\,,  $s=-1$
\begin{equation}
M^{\rm LCP}_{-1\,{\bf k}}=\,-i\,A_x\sin(\theta_{\bf
q}/2)+\,A_y\cos(\theta_{\bf q}/2)\,, \label{eq:mtx2_mono_2}
\end{equation}
and
\begin{equation}
M^{\rm RCP}_{-1\,{\bf k}}=\,-i\,A_x\sin(\theta_{\bf
q}/2)-\,A_y\cos(\theta_{\bf q}/2)\,. \label{eq:mtx2_mono_2}
\end{equation}
It follows that
\begin{equation}
I_{+1\,{\bf k}}= A_x^2\cos^2(\theta_{\bf
q}/2)+A_y^2\sin^2(\theta_{\bf q}/2)\,, \label{eq:I_cp_1}
\end{equation}
and
\begin{equation}
I_{-1\,{\bf k}}= A_x^2\sin^2(\theta_{\bf
q}/2)+A_y^2\cos^2(\theta_{\bf q}/2)\,, \label{eq:I_cp_2}
\end{equation}
regardless of the chirality of the light. This simple algebra
indicates that, within the tight-binding formalism for $|{\bf q}|\ll
|{\bf K}|$, the photoelectron intensity of graphene does not show
circular dichroism. Moreover, for perfect circular polarization, for
which $|A_x|=|A_y|$, $I_{s\,{\bf k}}$ is isotropic around the
constant energy contour. However, in real measurements, the
photoelectron intensity for X- and Y-polarization geometries are
different due to nonzero polarization component along the $z$
direction, e.\,g.\,, $A_x$ exhibits a finite $\hat{z}$ component
when projected to the sample surface, whereas $A_y$ has negligible
out-of-plane component~\cite{Hwang}, when the photoelectron
intensity is closely related to the scattering probability of the
real $p_z$ orbitals for the out-of-plane component of light
polarization, which is beyond the capability of the tight-binding
approach. For example, at a photon energy of 50~eV, the ratio of
$I_{\bf k}^{\rm X-pol.}/I_{\bf k}^{\rm Y-pol.}$ is
$\sim$21.4~\cite{Hwang} that we have used throughout our study. This
ratio is not controllable, but determined for each experimental
setup for each photon energy. This is also applied to the circular
polarization, for which ${\bf A}=A_x\,\hat{x}\pm\,i\,A_y\,\hat{y}$
is projected onto the sample surface.

  \begin{figure}[b]
  \begin{center}
  \includegraphics[width=0.95\columnwidth]{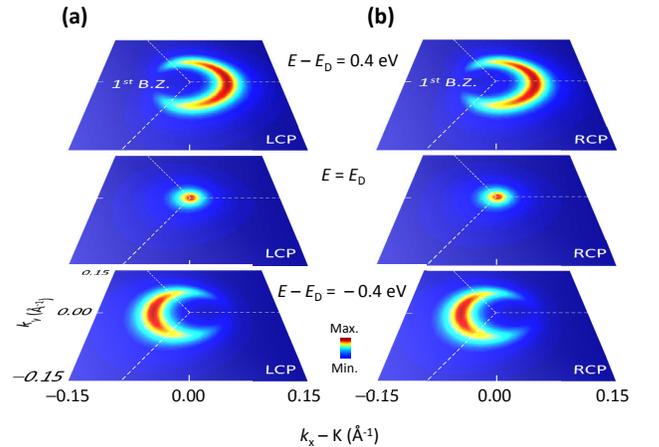}
  \end{center}
  \caption{(a, b) Calculated photoelectron intensity of graphene for (a) left- and (b) right-
  circularly polarized lights. An arbitrary energy
broadening of 0.10~eV has been used to qualitatively compare with
experimental results~\cite{Liu,Gierz}.}
  \label{Fig2}
  \end{figure}

These theoretical results are summarized in Fig.~2 at several
different energies with respect to the Dirac energy, $E_{\rm D}$.
For LCP (Fig.~2(a)), the constant energy contour shows a
crescent-like shape with minimum intensity in the first Brillouin
zone (1st B.Z.) at $E-E_{\rm D}=0.4$~eV and a point-like constant
energy map at $E_{\rm D}$. The intensity distribution is reversed at
$E-E_{\rm D}=-0.4$~eV, with respect to $E_{\rm D}$, with maximum
intensity in the 1st B.Z. While the overall shape shows the
characteristic conical dispersion of graphene, the intensity
distribution is similar to the case of
X-polarization~\cite{Hwang,Liu}. It is important to note that, for
linear polarizations, the crescent-like shape is understood by the
interference between photoelectrons emitted from two carbon
sublattices~\cite{Himpsel}. On the other hand, for circular
polarizations, it originates from the relatively weak photoelectron
intensity for Y-polarization as discussed above. As a result, the
difference in photoelectron intensity between LCP and RCP is not
determined by the properties of initial states, but given by the
experimental geometry, and hence we expect the same intensity
distribution for RCP as shown in Fig.~2(b).

Note also that our results (Eqs.~(14-15) and~(18)) for the
conduction band are different from those in the previous study
(Eq.~(3) of the Ref.~\cite{Liu}):
\begin{equation}
I= \left| A_x \xi_x (e^{i\theta/2}+e^{-i\theta/2}) \pm i\,A_y \xi_y
(e^{i\theta/2}-e^{-i\theta/2})\right|^2\,, \label{eq:Liu_1}
\end{equation}
where $+$ and $-$ signs correspond to LCP and RCP, respectively. In
this study~\cite{Liu}, they have claimed that the dipole transition
matrix elements $\xi_x$ and $\xi_y$ for the {\it x} and {\it y}
components of the vector potential have a relation of
$\xi_x\approx\xi_y$ at 30~eV~\cite{comment}. This setup results in
following photoelectron intensity (Eq.~(4) of Ref.~\cite{Liu}) for
the perfect circular polarization, i.\,e.\,, $|A_x|=|A_y|\equiv|A|$,
\begin{eqnarray}
I=4\left|\xi_x\right|^2A^2\left|\cos(\theta/2)\pm\sin(\theta/2)\right|^2.
\label{eq:Liu_2}
\end{eqnarray}
The difference between the two theoretical approaches
(Eq.~(\ref{eq:I_cp_1}) vs. Eq.~(\ref{eq:Liu_2})) originates from the
evolution of the tight-binding Hamiltonian. For example, for $|{\bf
q}|\ll |{\bf K}|$, the Hamiltonian is expressed via Pauli matrices,
i.\,e.\,, $H=v_{\rm F}\vec{\sigma}\cdot\vec{q}$ (Eq.~(8)), where
$v_{\rm F}$ is the Fermi velocity, with which we can obtain the
tight-binding eigenstates (used in both works done by us (Eq.~(10))
and Liu {\it et al.}~\cite{Liu}). Hence, the Peierls substitution
naturally leads to $H^{\rm
int}=\vec{\sigma}\cdot\vec{A}$~\cite{Neto}, both of which components
exhibit a strong influence on the photoemission matrix element when
applied to the spinor eigenstate of graphene~\cite{Hwang} as we have
done in our study. On the other hand, Liu {\it et al.}~\cite{Liu},
have applied the local Hamiltonian, $H={\bf p}^2/2m + V({\bf r})$,
to the tight-binding eigenstates (Eq.~(10)) obtained by the nonlocal
tight-binding Hamiltonian (Eq.~(1))~\cite{Louie}. Consequently, the
{\it y} component of the matrix element differs by the imaginary
number ``$i$'' arising from $\sigma_y$ (compare Eqs.~(14-15)
and~(\ref{eq:Liu_2})) resulting in the completely different
photoelectron intensity distributions [this issue is eliminated when
the realistic eigenstates, e.\,g.\,, maximally localized Wannier
functions obtained by {\it ab initio} calculations~\cite{Marzari},
are used in conjunction with the local Hamiltonian].

In real measurements, in contrast to our prediction, the
photoelectron intensity from graphene exhibits circular
dichroism~\cite{Liu,Gierz,Hwang_JPCM}. The intensity maximum around
a constant energy contour rotates by 180~degrees upon changing the
chirality of light with an energy of 30~eV~\cite{Liu}. Such a
dichroic effect varies at different photon energies~\cite{Gierz},
which has been attributed to the symmetry of the final states
($d$-like partial waves above a photon energy of 52~eV and $s$ or
$p$-like partial waves below 52~eV~\cite{Gierz_PRB}). The photon
energy dependence suggests that, when the final state effect is
minimized around 52~eV, the still observed circular dichroism
originates from the psuedospin nature of charge carriers in
graphene~\cite{Gierz}. Another study shows that, at a similar photon
energy of 50~eV, the circular dichroism changes as a function of
electron binding energy, which suggests a possibility of many-body
interactions as an origin of the observed
dichroism~\cite{Hwang_JPCM}. These ambiguities in understanding
experimental results may suggest the existence of an extrinsic
factor contributing to the circular dichroism that changes at
different experimental geometries and photon energies, such as
imperfect circular polarization. In fact, the polarization purity
for LCP and RCP in the experimental work for the Berry phase
scenario is $\sim$80~\%~\cite{Liu}.

  \begin{figure}[t]
  \begin{center}
  \includegraphics[width=0.95\columnwidth]{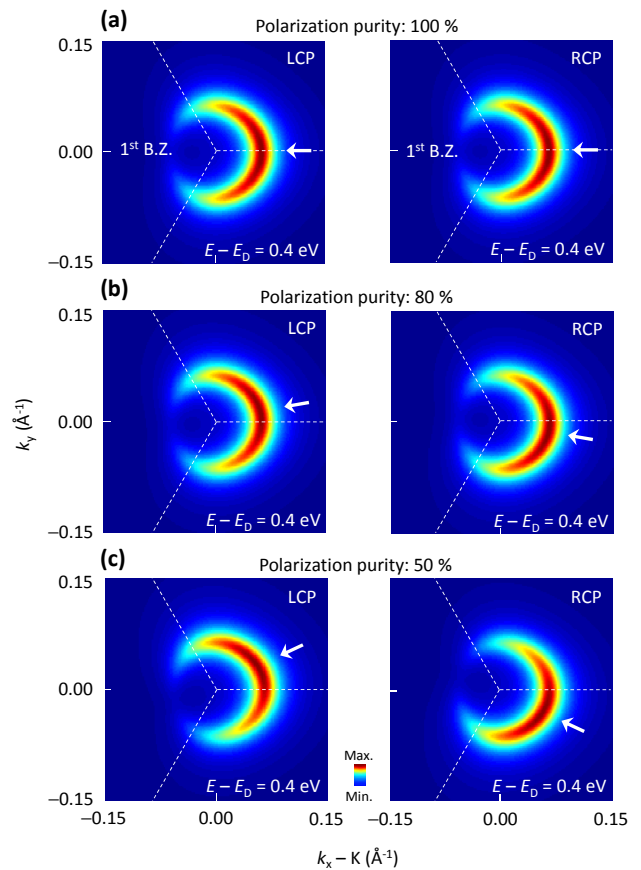}
  \end{center}
  \caption{(a-c) Calculated photoelectron intensity of graphene for LCP (left panels)
  and RCP (right panels) at $E-E_{\rm D}=0.4$~eV for 100~\%, 80~\%, and 50~\% circular polarization. An arbitrary energy
broadening of 0.10~eV has been used. The photon energy is assumed to
be 50~eV and the angle of incident photons and surface normal is 26
degrees, adapting the experimental geometry in Ref.~\cite{Hwang}.}
  \label{Fig3}
  \end{figure}

In Fig.~3, we show the effect of the imperfect polarization on the
rotation of intensity maximum upon changing the chirality of light.
We assume that the angle between the photon incident and the
electron detector angles is 55 degrees (e.\,g.\,, the experimental
setup for the previous experiment~\cite{Hwang}). Then the angle
between the photon incident and the sample normal angles is
determined by photon energies. For example, at 50~eV, where the {\bf
K} points is tilted by $\sim$29 degrees with respect to the sample
normal, the angle between the photon incident and the sample normal
angles is $\sim$26 degrees (14 and 31~degrees for 30 and 70~eV
photons, respectively). In addition, for imperfect circular
polarization, the major axis of the elliptical polarization is
tilted by $\pm$45 degrees for LCP and RCP, respectively, with
respect to $+\hat{x}$ direction. As a result, the elliptical
polarization $A_x\,\hat{x}\pm\,i\,A_y\,\hat{y}$ projected to the
sample surface becomes
$A_x\cos{26^{\circ}}\cos{45^{\circ}}\pm\,i\,A_y\cos{26^{\circ}}\sin{45^{\circ}}$
and $-A_x\sin{45^{\circ}}\pm\,i\,A_y\cos{45^{\circ}}$ along {\it x}-
and {\it y}-axis, respectively, where $(1-(a-b)/(a+b))\times100~\%$
defines the purity of the polarization. When the purity of the
polarization is 100~\% (Fig.~3(a)), the intensity maximum around the
constant energy contour at $E-E_{\rm D}=0.4$~eV stays the same upon
changing the chirality of light as denoted by the white arrows. When
the purity of the polarization decreases, the intensity maximum
becomes separated as denoted by the white arrows in Fig.~3(b) giving
rise to the circular dichroism. The circular dichroism becomes
stronger with further decreasing purity of the polarization as shown
in Fig.~3(c).

To further discuss the discrepancy of our results compared to
experimental results, we directly compare the rotational angles in
both our and previous studies. When the final state effect is
minimal at 52~eV~\cite{Gierz_PRB}, i.\,e.\,, intrinsic properties of
graphene dominate photoelectron intensities, the rotation of
intensity maxima is only $\sim$40 degrees~\cite{Gierz}, different
from the rotation by 180~degrees corresponding to the Berry phase
scenario~\cite{Liu}. While the tight-binding approach that we have
used excellently reproduces the experimental data for linearly
polarized light with an energy of 50~eV~\cite{Hwang}, the observed
($\sim$40~degrees) and predicted ($\sim$60~degrees via the first
principles calculations) dichroism~\cite{Gierz} is comparable to our
result of 40-60~degrees for the polarization purity of 80-50~\%
(Fig.~3).

Now let us restrict our discussion to the case where the photon
energy only changes photon polarization with respect to the sample
surface. With this set up, the simulations for left-circularly
polarized light with a polarization vector corresponding several
photon energies (30, 50, and 70~eV in Figs.~4(a-c), respectively).
Here, the angle between incident photons and the analyzer is assumed
to be 55~degrees. This simulation does not show a notable rotation
of photoelectron intensity around a constant energy contour upon
changing photon energy, suggesting that the final state
effect~\cite{Gierz} plays a dominant role in reproducing the photon
energy dependence, while the experimental condition in conjunction
with intrinsic effects, such as the pseudopin nature of charge
carriers~\cite{Gierz} and the many-body effects~\cite{Hwang_JPCM},
also plays a finite role in rotating photoelectron intensity upon
changing the chirality of light at constant photon energy. Overall,
the theoretical approach within tight-binding formalism predicts an
additional factor for the circular dichroism and hence suggests the
importance of experimental conditions to understand circular
dichroism observed via photoemission spectroscopy.

  \begin{figure}[h]
  \begin{center}
  \includegraphics[width=1\columnwidth]{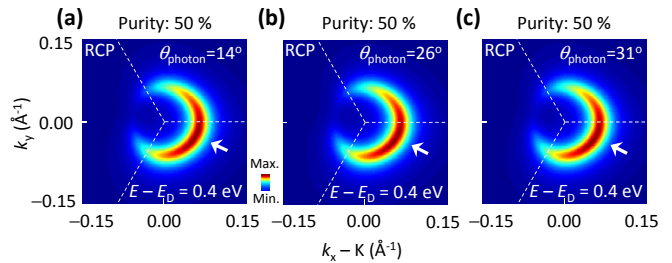}
  \end{center}
  \caption{(a-c)
Calculated photoelectron intensity of graphene for RCP at $E-E_{\rm
D}=0.4$~eV. An arbitrary energy broadening of 0.10~eV has been used.
The photon energy is assumed to be (a) 30, (b) 50, and (c) 70~eV,
where the angle of incident photons and surface normal is 14, 26,
and 31~degrees, adapting the experimental geometry in
Ref.~\cite{Hwang}.}
  \label{Fig4}
  \end{figure}

\section{Conclusion}
We have investigated the role of imperfect circular polarization in
the circular dichroism of photoelectron intensity observed in
graphene. Within the tight-binding formalism, we found that the
calculated photoemission matrix element does not predict any
difference between left- and right-circular polarization. However,
we found that circular dichroism is developed and enhanced with
decreasing purity of the polarization. Our results implies that the
experimental conditions should be taken into account to understand
circular dichroism, which invites further experimental and
theoretical investigation to understand the origin of the observed
circular dichroism in graphene.

\acknowledgments We are greatly indebted to Cheol-Hwan Park and
Sung-Kwan Mo for great help that has made this study possible. This
work was supported by a 2-Year Research Grant of Pusan National
University.

\end{document}